\documentclass[aps,prl,reprint,nofootinbib]{revtex4-1}

\usepackage[utf8]{inputenc}
\usepackage{color}
\usepackage{amsmath,amssymb}
\usepackage{amsfonts}
\usepackage{verbatim}
\usepackage{makeidx}
\usepackage{hyperref}
\usepackage{slashed}
\usepackage[vcentermath]{youngtab}
\usepackage{float}
\usepackage{booktabs}
\usepackage{graphicx}
\usepackage{mathrsfs}
\usepackage{bm}
\usepackage{braket}
\usepackage{url}
\usepackage{etoolbox}
\apptocmd{\sloppy}{\hbadness 10000\relax}{}{}
\allowdisplaybreaks[3]
\def\arraystretch{1.0}

\newcommand{\be}{\begin{eqnarray}}
\newcommand{\ee}{\end{eqnarray}}

\usepackage[normalem]{ulem}  
\renewcommand\sout{\bgroup \color{red} \ULdepth=-.5ex \ULset}

\begin{document}

\title{$P_c$ pentaquarks  with chiral tensor and quark dynamics}

\author{Yasuhiro~Yamaguchi$^{1}$}
\email{yasuhiro.yamaguchi@riken.jp}
\author{Hugo~Garc\'ia-Tecocoatzi$^{2}$}
\author{Alessandro~Giachino$^{3,4}$}
\author{Atsushi~Hosaka$^{5,6}$}
\author{Elena~Santopinto$^{3}$}
\email{elena.santopinto@ge.infn.it}
\author{Sachiko~Takeuchi$^{7,1,5}$}
\author{Makoto~Takizawa$^{8,1,9}$}
\email{takizawa@ac.shoyaku.ac.jp}

\affiliation{$^1$
Theoretical Research Division, Nishina Center, RIKEN, Hirosawa, Wako, Saitama 351-0198, Japan
}
\affiliation{$^2$Instituto de Ciencias Nucleares, Universidad Nacional Aut\'onoma de M\'exico, 
AP 70-543, 04510 M\'exico DF, M\'exico
}
\affiliation{$^3$Istituto Nazionale di Fisica Nucleare (INFN), Sezione di
Genova, via Dodecaneso 33, 16146 Genova, Italy
}
\affiliation{$^4$Dipartimento di Fisica dell'Universit\`a di Genova, via Dodecaneso 33, 16146 Genova, Italy
}
\affiliation{$^5$Research Center for Nuclear Physics (RCNP), Osaka
University, Ibaraki, Osaka 567-0047, Japan}
\affiliation{$^6$Advanced Science Research Center, Japan Atomic Energy Agency, Tokai, Ibaraki 319-1195, Japan}
\affiliation{$^7$Japan College of Social Work, Kiyose, Tokyo 204-8555, Japan}
\affiliation{$^8$Showa Pharmaceutical University, Machida, Tokyo
194-8543, Japan}
\affiliation{$^9$J-PARC Branch, KEK Theory Center, Institute for Particle and Nuclear Studies, KEK, Tokai, Ibaraki 319-1106, Japan}
\date{\today}

\begin{abstract}
We investigate the hidden-charm pentaquarks as superpositions of $\Lambda_c \bar{D}^{(*)}$ and $\Sigma_c^{(*)} \bar{D}^{(*)}$  (isospin $I = 1/2$) meson-baryon channels coupled to a $uudc\bar{c}$ compact core by employing an interaction  satisfying the heavy quark and chiral symmetries.
Our model can 
 consistently explain the masses and decay widths of  $P_c^+(4312)$, $P_c^+(4440)$ and $P_c^+(4457)$ 
with the dominant components of $\Sigma_c \bar D$ and $\Sigma_c \bar D^\ast$ with spin parity assignments $J^P = 1/2^{-}, 3/2^{-}$ and $1/2^{-}$, respectively.  
We analyze basic properties of the $P_c$'s such as masses and decay widths, and find that the mass ordering is dominantly determined by the quark dynamics while the decay widths by the tensor force of the one-pion exchange.  

\end{abstract}

  \maketitle

  %
   In 2015, the Large Hadron Collider beauty experiment (LHCb) collaboration observed two hidden-charm
  pentaquarks, $P^+_c(4380)$ and $P^+_c(4450)$, in
  $\Lambda^0_b\rightarrow J/\psi K^-p$ decay  ~\cite{Aaij:2015tga} and reported additional analysis efforts \cite{Aaij:2016phn,Aaij:2016ymb}. 
 These results have motived  hundreds of theoretical  articles (just to make some examples see 
 \cite{Chen:2016qju,Ali:2017jda,Yuan:2012wz,Takeuchi:2016ejt,Santopinto:2016pkp,Wu:2010jy,Wu:2010vk,Garcia-Recio:2013gaa,Karliner:2015ina,Chen:2015loa,Roca:2015dva,He:2015cea,Meissner:2015mza,Chen:2015moa,Uchino:2015uha,Burns:2015dwa,Yamaguchi:2016ote,Shimizu:2017xrg,Kubarovsky:2015aaa,Huang:2013mua,Huang:2016tcr,Garzon:2015zva,Liu:2016dli,Kim:2016cxr,Wang:2015jsa,Shimizu:2016rrd,Wu:2010rv,Xiao:2013jla,Azizi:2017bgs,Cheng:2016ddp}).
Recently a new analysis has been reported \cite{Aaij:2019vzc} using nine times more data from the Large
Hadron Collider than the 2015 analysis. The data set
was first analyzed in the same way as before and the parameters of the previously reported
$P^+_c(4450)$, and  $P^+_c (4380)$ structures were consistent with the original results. As well as revealing
the new $P^+_c(4312) $ state, the analysis also uncovered a more complex structure of $P^+_c(4450)$, 
consisting of two narrow nearby separate peaks, $P^+_c (4440)$ and $P^+_c(4457)$, with the two-peak structure hypothesis
having a statistical significance of 5.4 sigma with respect to the single-peak structure hypothesis. 
As for a broad state $P^+_c (4380)$ (width $\sim 200$MeV), in the new analysis using higher-order polynomial 
functions for the background, data can be fitted equally well without the Breit-Wigner contribution corresponding to broad $P^+_c (4380)$ state.  
In this situation, more experimental and theoretical studies are
needed to fully understand the structure of the observed states.

The masses and widths of the three narrow pentaquark states are as follows \cite{Aaij:2019vzc}.
 \begin{eqnarray}
\nonumber     P_c^+(4312):     M &=& 4311.9 \pm  0.7 ^{+6.8}_{-0.6}  \mbox{ MeV} \, ,
\\ \nonumber            \Gamma &=&    9.8 \pm  2.7 ^{+3.7}_{-4.5}  \mbox{ MeV} \,;
\\ \nonumber  P_c^+(4440):     M &=& 4440.3 \pm  1.3 ^{+4.1}_{-4.7}  \mbox{ MeV} \, ,
\\ \nonumber            \Gamma &=&   20.6 \pm  4.9 ^{+8.7}_{-10.1} \mbox{ MeV} \, ;
\\ \nonumber  P_c^+(4457):     M &=& 4457.3 \pm  0.6 ^{+4.1}_{-1.7}  \mbox{ MeV} \, ,
\\   \nonumber                    \Gamma &=&    6.4 \pm  2.0 ^{+5.7}_{-1.9}  \mbox{ MeV} \, .
\label{experiment}
\end{eqnarray}
As discussed by LHCb~\cite{Aaij:2019vzc}, $P_c^{+}(4312)$ is just below the $\Sigma_c \bar D$ 
threshold, while the higher ones $P_c^+(4440)$ and $P_c^{+}(4457)$ are both below the $\Sigma_c \bar D^{*}$ threshold.
This change of the experimental observation motivated 
new theoretical investigations \cite{Chen:2019bip,Xiao:2019aya,Liu:2019tjn,Shimizu:2019ptd,Giannuzzi:2019esi,Giron:2019bcs,He:2019ify,Ali:2019npk,Guo:2019kdc,Burns:2019iih}. Among them, \cite{Chen:2019bip,Xiao:2019aya,He:2019ify,Burns:2019iih} are taking the hadronic molecule approach.
In \cite{Chen:2019bip} 
the authors explore several scenarios for the structures of the pentaquark states
by means of QCD sum rules. They propose to interpret all the
four  pentaquarks as molecular states, in particular they interpret $P_c^+(4440)$ and $P_c^+(4457)$ as 
$\Sigma_c^{\ast ++} \bar D^{-}$ and $\Sigma_c^{+} \bar D^{\ast 0}$ molecular states, both with $J^P = 3/2^-$.
In \cite{Xiao:2019aya} pentaquark states are studied
with a local hidden gauge based interaction in a coupled channel approach by including
the $N\eta_c$, $NJ/\psi$, $\Lambda_c\bar{D}^{(*)} $ and $\Sigma_c^{(*)} \bar{D}^{(*)}$ meson-baryon channels. 
They assign $P_c^+(4440)$ to $J^P=1/2^{-}$ and $P_c^+(4457)$ to $J^P=3/2^{-}$.
Although these assignments agree with experimental decay widths,
the mass of $P_c^+(4440)$ is overestimated by about 13 MeV, which is more than the double of the 
experimental error on the $P_c^+(4440)$ mass of about 5 MeV.
Most importantly, the mass difference between $P_c^+(4440)$ and $P_c^+(4457)$, approximately $17$ MeV,
is not reproduced by this model in which, instead, the two states are almost degenerate.
$P_c^+(4440)$ and $P_c^+(4457)$ are considered as the $\Sigma_c^{(\ast)} \bar D^{(\ast)}$ hadronic molecule states in 
a quasipotential Bethe-Salpeter equation approach \cite{He:2019ify}. They use the meson-exchange interaction with 
$\pi$, $\eta$, $\rho$, $\omega$ and $\sigma$ mesons and reproduce the observed masses reasonably. 
Their spin-parity assignments are $P_c^+(4440)$ as $1/2^-$ and $P_c^+(4457)$ as $3/2^-$, respectively. 
Coupled-channel molecular states of 
the relative $S$-$D$($P$)-wave $\Sigma_c\bar D{}^*$ and
the relative $P$($S$-$D$)-wave $\Lambda_c(2595)\bar D$
are studied with OPEP in~\cite{Burns:2019iih}
as the $\Lambda_c(2595)\bar D$ threshold is very closed to the
$P^+_c(4457)$ mass.
The model predicts two bound states,
which they argue correspond to $P^+_c(4440)$($J^P=3/2^-$)
and $P^+_c(4457)$($J^P=1/2^+$).

In Ref.~\cite{Yamaguchi:2017zmn} we studied the hidden-charm pentaquarks by coupling the $\Lambda_c \bar{D}^{(*)}$ and $\Sigma_c^{(*)} \bar{D}^{(*)}$ meson-baryon channels  to a $uudc\bar{c}$ compact core with a meson-baryon binding interaction satisfying 
the heavy quark and chiral symmetries.
In that work we expressed the hidden-charm pentaquark masses and decay widths as functions of one free parameter, 
which is proportional to the coupling  strength between the meson-baryon and 5-quark-core states.
Interestingly enough, we find that the model has predicted the  masses and decay widths consistently with the new data with the following quantum number assignments: 
$J^P_{P_c^+(4312)^{}} = 1/2^{-}$, $J^P_{P_c^+(4440)^{}} = 3/2^{-}$ and $J^P_{P_c^+(4457)^{}}= 1/2^{-}$.
Our assignments of the quantum numbers for the $P_c^+(4440)$ and $P_c^+(4457)$ states are different from those 
in other hadronic-molecule approaches.

The purpose of the present article is to study the origin of the mass difference between $P_c^+(4440)$ and $P_c^+(4457)$ by
performing the calculations with and without the tensor term of the one-pion exchange potential (OPEP). 
The importance of the tensor force is emphasized as ``chiral tensor dynamics''.

Let us briefly overview the main ingredients of the model of Ref. \cite{Yamaguchi:2017zmn}. 
The best established interaction between the meson and the baryon is provided
by OPEP, which is obtained by the effective Lagrangians 
satisfying the heavy quark and chiral symmetries.
The interaction Lagrangian between the 
ground state heavy mesons, $\bar{D}$ and $\bar{D}^\ast_{}$, and the pions can be written
in a compact form \cite{Wise:1992hn,Falk:1992cx,Casalbuoni:1996pg,Manohar:2000dt}, 
\be
 {\cal L}_{\pi HH}=g_A^M {\rm Tr}\left[H_b \gamma_\mu\gamma_5 A^{\mu}_{ba}\bar{H}_{a}\right],
\ee
where $H_a =  [ \bar D^\ast_{a\mu} \gamma^\mu - \bar D_a \gamma_5 ](1 + \gamma_\mu v^\mu)/2$ and 
$\bar{H}_a = \gamma_0 H^\dagger_a \gamma_0$ are the heavy meson fields containing the spin multiplet of pseudoscalar and 
vector meson fields $\bar D_a$ and $\bar D^\ast_{a\mu}$.
The trace ${\rm Tr}\left[\cdots\right]$ is taken over the gamma matrices.
The subscript $a$ denotes the light quark flavor, and $v_{\mu}$ is the four-velocity of the heavy quark inside the heavy meson;
 $g_A^M$  is the the axial vector coupling constant for heavy mesons, 
 which was determined by the $D^{*}\to D \pi$ strong decay to be $g_A^M = 0.59$ 
 \cite{Casalbuoni:1996pg,Manohar:2000dt,Tanabashi:2018oca}, and 
$A^{\mu}_{}=\frac{i}{2}\left[\xi^\dagger \partial_\mu
 \xi - \xi \partial_\mu \xi^\dagger \right]$, with $\xi = \exp\left(\frac{i\hat{\pi}}{2f_\pi}\right)$,  is the pion axial vector current;
 $\hat{\pi}$ is  the  flavor matrix of the pion field and $f_\pi=92.3$  MeV 
 is the pion decay constant.
The effective Lagrangian which describes the interaction between  $\Sigma_c$ and
 $\Lambda_c$  heavy baryons and the pions is~\cite{Yan:1992gz,Liu:2011xc}
\be
 {\cal L}_{\pi
 BB} & = & \frac{3}{2}g_1(iv_\kappa)\varepsilon^{\mu\nu\lambda\kappa}{\rm tr}\left[
 \bar{S_\mu}A_\nu S_\lambda\right]
 + g_4{\rm tr}\left[\bar{S}^\mu A_{\mu} \hat{\Lambda}_{\rm c} \right]   \nonumber \\
 & & + {\rm H.c.},
\ee
where  ${\rm tr}\left[\cdots\right]$ denotes the trace performed in  flavor space.
The superfields $S_\mu$ and $\bar{S}_\mu$ are represented by
\be
 S_{\mu}=
\hat{\Sigma}^{\ast}_{{\rm c}\mu}
 -\frac{1}{\sqrt{3}}\left(\gamma_\mu+v_\mu\right)\gamma_5
 \hat{\Sigma}_{\rm c}, \, \, 
 \bar{S}_\mu = S^\dagger_\mu \gamma_0.
\ee
Here, the heavy baryon fields
$\hat{\Lambda}_{\rm c}$ and $\hat{\Sigma}^{(\ast)}_{\rm c (\mu)}$, 
are 
\be
 \hat{\Lambda}_{\rm c}
 =
\left(
 \begin{array}{cc}
  0&\Lambda^+_{\rm c}  \\
  -\Lambda^+_{\rm c} & 0  \\
 \end{array}
 \right), 
\ee 
\be 
 \hat{\Sigma}^{(\ast)}_{\rm c (\mu)}
 =\left(
 \begin{array}{cc}
  \Sigma^{(\ast)++}_{\rm c (\mu)}&\frac{1}{\sqrt{2}}\Sigma^{(\ast)+}_{\rm c (\mu)}  \\
  \frac{1}{\sqrt{2}}\Sigma^{(\ast)+}_{\rm c (\mu)} &\Sigma^{(\ast)0}_{\rm c (\mu)}  \\
 \end{array}
 \right) .
\ee
As shown in ~\cite{Liu:2011xc}, $g_1=(\sqrt{8}/3)g_4=1$. 
The internal structure of hadrons is parametrized by a dipole form factor at each vertex,
$
 F(\Lambda,\bm{q}\,)=
 \frac{\Lambda^2-m^2_\pi}{\Lambda^2+\bm{q}\,^2}
$,
where  $m_\pi$ and $\bm{q}$ are the mass and three-momentum of an
incoming pion and  the heavy hadron cut-offs  $\Lambda_{H}$ are determined by the ratio between
the sizes of the heavy hadron, $r_{H}$, and the nucleon, $r_N$, 
$
\Lambda_N/\Lambda_{H}=r_{H}/r_N
$,
We obtained $\Lambda_{\Lambda_{\rm c}} \sim \Lambda_{\Sigma_{\rm c}} \sim \Lambda_N$ for the charmed baryons and  $\Lambda_{\bar{D}}=1.35\Lambda_N$  for the $\bar{D}^{(\ast)}$ meson, where
the nucleon cutoff is determined to reproduce the deuteron-binding
energy by the one-pion exchange potential(OPEP) as $\Lambda_N=837$ MeV~\cite{Yasui:2009bz,Yamaguchi:2011xb,Yamaguchi:2011qw}.
The explicit form of the OPEP $V^\pi (\bm{q})$ between the meson-baryon ($MB$) channels in the momentum space is as follows,
\be
 V^\pi(\bm{q})=-\left(\frac{g_A^M g_A^B}{4f_\pi^2} \right)
 \frac{(\bm{\hat{S}}_1\cdot \bm{q}\,)(\bm{\hat{S}}_2\cdot \bm{q}\,)}{\bm{q}\,^2+m^2_{\pi}}\bm{\hat{T}}_1\cdot \bm{\hat{T}}_2 \, ,
 \label{eq_OPEP_MB}
\ee 
where $\bm{\hat{S}}$ is the spin operator and $\bm{\hat{T}}$ is the isospin operator. $g_A^B$ is the axial vector coupling constant of the corresponding baryons.~\footnote{
In our previous publication~\cite{Yamaguchi:2016ote}, 
there were a few errors in the matrix elements, which are corrected in this paper.
After the corrections, however, important results of our discussions remain unchanged. 
}

The coupling of the $MB$ channels, $i$ and $j$, to the five-quark ($5q$) channels, $\alpha$,
gives rise to an effective interaction, $V^{5q}$, 
\be
   \Braket{ i | V^{5q} | j}  = \sum_\alpha \Braket{ i | V | \alpha} \frac{1}{E- E^{5q}_\alpha} \Braket{ \alpha | V^\dagger | j}  \, ,
   \label{eq:approximationV}
\ee
where $V$ represents the transitions between the $MB$ and $5q$ channels and $E^{5q}_\alpha$ is the eigenenergy of a $5q$ channel.
We further introduced the following assumption,
\be
  \Braket{ i | V | \alpha}  = f \Braket{i | \alpha} \, ,
  \label{eq:approximationV2} 
\ee
where $f$ is the only free parameter which determines the overall strength of the matrix elements.
In order to calculate the $\Braket{i | \alpha}$, we construct the meson-baryon and five-quark wave functions
explicitly in the standard non-relativistic quark model with a harmonic oscillator confining potential.
The derived potential $\Braket{ i | V^{5q} | j}$ turned out to give similar results to those derived from the quark cluster model \cite{Takeuchi:2016ejt}.

The energies and widths of the bound and resonant states were obtained by solving the 
coupled-channel Schr\"odinger equation with the OPEP, $V^\pi(\bm{r})$, and $5q$ potential $V^{5q}(\bm{r})$,
\be
\left( K + V^\pi(\bm{r}) + V^{5q}(\bm{r}) \right) \Psi(\bm{r}) = E \Psi(\bm{r}) \, ,
\label{eq:Schr}
\ee
where $K$ is the kinetic energy of the meson-baryon system and $\Psi(\bm{r})$ is the wave function of the meson-baryon systems 
with $\bm{r}$ being the relative distance between the center of mass of the meson and that of the baryon.
The coupled channels included are all possible ones of $\Sigma_c^{(\ast)} \bar D^{(\ast)}$ and 
$\Lambda_c \bar D^{(\ast)}$ which can form a given $J^P$ and isospin $I = 1/2$.

Eq.~(\ref{eq:Schr}) is solved by using 
variational method. We used the Gaussian basis functions as trial functions \cite{Hiyama:2003cu}.  In order to obtain 
resonance states, we employed the complex scaling method \cite{moiseyev_2011}.

\begin{figure}
\includegraphics[width=0.8\linewidth]{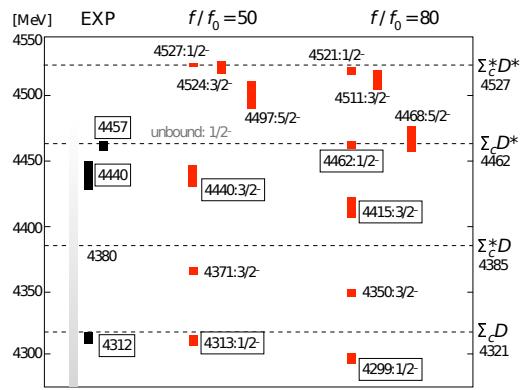}
  \caption{
 (Color online)
 Experimental data (EXP)~\cite{Aaij:2015tga,Aaij:2019vzc} and our 
results of
masses and widths for various $P_c$ states. The horizontal dashed lines show the thresholds for corresponding channels and values in the right axis are isospin averaged ones in units of MeV.
The centers of the bars are located at the central values of pentaquark masses while their lengths correspond to 
the pentaquark widths with the exception of $P_c(4380)$ width.
 }
  \label{comparisonNew}
\end{figure}

In Fig.~\ref{comparisonNew} and Table~\ref{tab:result01}, 
experimental data~\cite{Aaij:2015tga,Aaij:2019vzc} and our predictions are compared.  
The centers of the bars in Fig.~\ref{comparisonNew} are located at the central values of pentaquark masses while their lengths correspond to 
the pentaquark widths with the exception of $P_c(4380)$ width, which is too large and does not fit into the shown energy region.  
The boxed numbers are the masses of the recently observed states~\cite{Aaij:2019vzc}, and the corresponding 
predictions in our model.
The dashed lines are for threshold values.  
Our predicted masses and the decay widths are 
shown for the parameters $f/f_0=50$ and $f/f_0=80$.
Here, $f_0$ is the strength of the one-pion exchange diagonal term for the $\Sigma_{\rm c} \bar{D}^\ast$ meson-baryon channel, 
$
f_0=
 \left|C^{\pi}_{\Sigma_{\rm c}\bar{D}^{\ast}}(r=0)\right|\sim 6 \text{ MeV}
$ 
(see Ref.~\cite{Yamaguchi:2017zmn}).
Setting the free parameter $f/f_0$ at $f/f_0=50$,
we observe that  both masses and widths of 
$P_c^+(4312)$ and $P_c^+(4440)$
are reproduced within the experimental errors.
However, the state corresponding to $P_c^+(4457)$ is absent 
in our results,
where the attraction is not enough.
Increasing the value of $f/f_0$ to 70,
the state with $J^P=1/2^-$ appears below the $\Sigma_c\bar{D}^\ast$ threshold, and at $f/f_0=80$ the mass and width of this state are in reasonable agreement with $P_c^+(4457)$.
However, as shown in Fig.~\ref{comparisonNew},
the attraction at $f/f_0=80$ is 
stronger than that at $f/f_0=50$ and hence
the masses of the other states shift downward.

We find as expected that the dominant components of these states are nearby threshold channels and with the quantum numbers as follows; 
 $\Sigma_c^{} \bar D^{}$ with $J^P=1/2^{-}$ ($P_c^{+}(4312)$), $\Sigma_c^{} \bar D^{*}$ with
 $J^P=3/2^{-}$  ($P_c^{+}(4440)$) and  with  $J^P=1/2^{-}$  ($P_c^{+}(4457)$) meson-baryon molecular states. 

%
%

Let us compare our 
results with the ones reported by other works.
In Ref.~\cite{Xiao:2019aya},  
the assignments of the quantum numbers for 
$P_c(4440)$ and $P_c(4457)$ are different from ours.
\renewcommand{\arraystretch}{1.2}
\begin{table*}
\begin{center}
\caption{
 Comparison between the experimental mass spectrum and decay widths with our  
 results.
 For our results for $f/f_0=80$, the values in parentheses are obtained without the OPEP tensor force, which are also shown in Fig.~\ref{fig:Tensor}.
All values except $J^P$ are in units of MeV.}
\begin{tabular}{ccc||ccc|ccc}
\toprule[1pt]\toprule[1pt]
\multicolumn{3}{c||}{EXP~\cite{Aaij:2015tga,Aaij:2019vzc}} & \multicolumn{3}{c|}{Our 
 results 
for $f/f_0=50$}& \multicolumn{3}{c}{Our results 
 for $f/f_0=80$} \\
~~\mbox{State}~~ & ~~\mbox{Mass} ~~ & ~~\mbox{Width}~~ & 
 ~~\mbox{$J^P$}~~ &  ~~\mbox{Mass} ~~ & ~~\mbox{Width}~~ &  
 ~~\mbox{$J^P$}~~ &  ~~\mbox{Mass} ~~ & ~~\mbox{Width}~~ \\ \midrule[1pt]
$P_c^{+}(4312)$        &$4311.9 \pm  0.7 ^{+6.8}_{-0.6} $   &  $9.8 \pm  2.7 ^{+3.7}_{-4.5}$   & 
 $1/2^-$ & 4313& 9.6 & 
 $1/2^-$ & 4299 (4307)& 9.4 (12) \\ \midrule[1pt]
$P_c^{+}(4380)$       &$4380 \pm 8 \pm29 $   &  $205 \pm 18 \pm 86 $   & 
 $3/2^-$ & 4371& 5.0 & 
 $3/2^-$ & 4350 (4365)& 5.0 (3.6) \\ \midrule[1pt]
 $P_c^{+}(4440)$        & $4440.3 \pm  1.3 ^{+4.1}_{-4.7}$   &  $20.6 \pm  4.9 ^{+8.7}_{-10.1}$   &
 $3/2^-$ & 4440& 16&
 $3/2^-$ & 4415 (4433)& 15 (1.8)\\ \midrule[1pt]
 $P_c^{+}(4457)$    &   $4457.3 \pm  0.6 ^{+4.1}_{-1.7} $ &   $ 6.4 \pm  2.0 ^{+5.7}_{-1.9} $    & 
 ---& ---& ---&
 $1/2^-$ & 4462 (4462)& 3.2 (0.96)\\   \midrule[1pt]
 &    &   &
 $1/2^-$ & 4527 & 0.88& 
 $1/2^{-}$ & 4521 (4526)& 2.8 (0.18)\\   \midrule[1pt]
&    &   &
 $3/2^-$	& 4524& 7.6 &
 $3/2^-$ & 4511 (4521)& 14 (3.4)\\   \midrule[1pt]
&    &   &
 $5/2^-$ & 4497 &20 &
 $5/2^-$ & 4468 (4491)&18 (0.0)\\ \bottomrule[1pt]
\end{tabular}
\label{tab:result01}
\end{center}
\end{table*}
\renewcommand{\arraystretch}{1.0}
Since these two states are located near $\Sigma_c \bar{D}^\ast$ threshold and both states have the narrow widths, 
it is natural to consider them to form the 
$J=1/2$ and $3/2$ states
in S-wave. 
It is emphasized that in our model the spin $3/2$ state (4440) is lighter than the spin $1/2$ state (4457).   
In Ref.~\cite{Liu:2019tjn}, they studied seven heavy quark multiplets of  $\Sigma_c\bar D$, $\Sigma_c\bar D^\ast$, $ \Sigma_c^\ast\bar D$, and $\Sigma_c^\ast\bar D^\ast$, 
and considered two options of inputs, $P_c(4440, 4457) \sim (3/2,1/2)$ which they call set A and (1/2, 3/2) set B.
In the heavy quark limit, there are two parameters in the Hamiltonian and so the above inputs for the two states are enough to fix the two parameters. The other five states are predicted.  
Interestingly, their set A predicts the other five states similarly to what our model predicts.  

Therefore, new LHCb results give us
an opportunity to study the spin-dependent forces between the $\Sigma_c$ and $\bar{D}^\ast$.
It is important to determine which of the above spin $1/2$ and $3/2$ states
is more deeply bound.  
There are two sources for the spin-dependent force in our model. 
One is the short range interaction by the coupling to the 5-quark-core states.
The other is the long range interaction by the OPEP, especially the tensor term.
%

\begin{figure}
\begin{center}
 \includegraphics[width=0.8\linewidth]{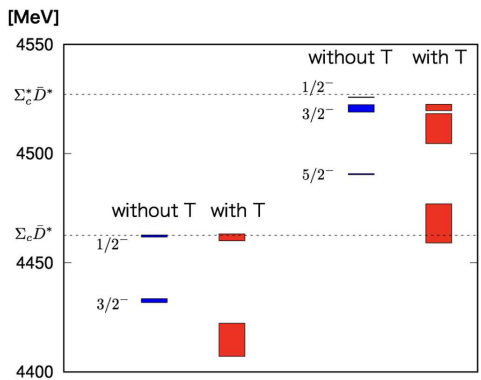}
\end{center}
  \caption{ \label{fig:Tensor}
 (Color online)
 Comparing the results with and without the tensor force of the OPEP
 for the states around the $\Sigma_c\bar{D}^\ast$ and 
 $\Sigma^\ast_c\bar{D}^\ast$ thresholds.
 The label 'without T'
 stands for the result without the OPEP tensor force, while 
 the label 'with T'
 stands that with the OPEP tensor force. 
The same convention is adopted as in Fig.~\ref{comparisonNew}.
 }  
\end{figure}%
To examine the effects of the tensor interaction of the OPEP, 
we have investigated the energy of the resonant $P_c$ states of 
$J=1/2$ and $3/2$ 
around the $\Sigma_c \bar D{}^*$ threshold,
and 
of $J=1/2$, 3/2 and 5/2 
around the $\Sigma_c^\ast \bar D{}^*$ threshold
without the OPEP tensor term as shown in Fig.~\ref{fig:Tensor}. 
In that plot, we have used $f/f_0=80$.
From Fig.~\ref{fig:Tensor}, we observe the following facts.
(1) The tensor force provides attraction as indicated by 
the results 
with T in Fig.~\ref{fig:Tensor}. This is because it contributes to the energy in the second order due to channel couplings.
(2) The role of the tensor force is further prominent in the decay width;
the agreement with the experimental data is significantly improved. 
Moreover, the decay width increases as the spin value increases.
We consider it again because of coupled-channel effects 
due to the OPEP tensor force. 
The dominant components of the obtained resonances are 
the $S$-wave state of the nearby threshold channel.
The tensor coupling allows the resonances to decay into the $D$-wave channels below the resonances.
Since there are many $D$-wave coupled channels in the higher spin states, 
the decay widths of these states are increased.
In fact,
the number of the $D$-wave coupled channels below the $\Sigma_c^\ast\bar{D}^\ast$ threshold is 3 for $J^P=1/2^-$, while 7 for $J^P=3/2^-, 5/2^-$.
From the observation in Fig.~\ref{fig:Tensor}, we find that
the short range 
interaction
is 
more attractive in the $3/2^-$ state in the present model.
This contrasts with what is 
expected for the color-spin interaction that provides more attraction for the 
$1/2^-$ state.  
The reason is in the quark structure of hadrons as explained below.  
In the quark cluster model, the hadron interaction 
is due mainly to the two terms: 
one is the Pauli-blocking effect which is measured by the norm (overlap) kernels
and the other is the color-spin interaction from the one gluon exchange.
The former is included in the present study, and is usually dominant when the norm of the two-hadron state 
deviates largely from 1~\cite{Shimizu:2000wm,Oka:2000wj}.
It can be less than 1 due to the Pauli-blocking (repulsive) but also can be more than 1 (attractive) 
because of the spectroscopic factor.
For the $\Sigma_c \bar D{}^*$ channel, the norm is ${23/18}$ for the $3/2^-$ state while $17/18$ for the 
$1/2^-$ state~\cite{Yamaguchi:2017zmn}.
Namely, this contribution of the spectroscopic factor 
is strongly attractive in the $\Sigma_c \bar D^\ast 3/2^-$ state and slightly repulsive in the $\Sigma_c \bar D^\ast 1/2^-$ state.

To estimate the effect of the color-spin interaction, which is not included  in the present study,
we revisit the coupled-channel 
dynamical calculation where both the Pauli-blocking and color-spin effects are 
included (but without the OPEP)~\cite{Takeuchi:2016ejt}.  
There a sharp cusp structure was observed for the $3/2^-$ state 
at  the $\Sigma_c \bar D{}^*$ threshold while not for $1/2^-$ state, 
implying some attraction for the  $3/2^-$ state while little for $1/2^-$ state.  
This is because the color-spin force which is attractive for the spin $1/2^-$ state 
has been overcome by the Pauli-blocking effect that acts reversed manner.  
Therefore, the effect of the color-spin force is not dominant for 
these resonant states.
It is interesting to investigate the system
with a model which includes both of the above quark model features with the OPEP,
which is now underway.
{
Although we have obtained the $J^P= 3/2^{-}$ state 
at $4371$ MeV and at $4350$ MeV 
when $f = 50$ and 80, respectively, 
we do not consider that 
this 
state corresponds to the LHCb's $P_c^+(4380)$ state. 
The observed $P_c^+(4380)$ has a width of about 200 MeV while 
that of our predicted state is only about $5$ MeV.  
In the first LHCb analysis \cite{Aaij:2015tga}, though the higher $P_c$ states were treated as one state, 
the opposite parity assignments were preferred for lower and higher $P_c$ states. On the other hand, 
all the states we have obtained here have the same parity minus.  
In the new LHCb analysis \cite{Aaij:2019vzc}, using higher-order polynomials for the background, 
data could be fitted without the broad $P_c^+(4380)$ Breit-Wigner resonance contribution.
Therefore, further theoretical as well as experimental studies are necessary for the $P_c^+(4380)$ state. 

}

In addition to the three states observed in the LHCb, we obtained four more states
including the one corresponding to $P_c^+(4380)$ near (below) the $\Sigma_c^\ast \bar D^{(\ast)}$ thresholds 
as shown in Fig.~\ref{comparisonNew} and Table~\ref{tab:result01}.
Due to the spins of $\Sigma_c^\ast$ and $\bar D^{(\ast)}$, $J = 3/2$ 
and 1 (or 0), respectively, they naturally form 
either 
triplet states, $J = 5/2, 3/2, 1/2$, 
or a singlet state of 
$J=3/2$.  
A possible reason that those states are not seen
would be due to a wider width of $\Sigma_c^*$.  
In fact, the width of the $\Sigma_c^\ast$ is about 15 MeV, 
while that of the $\Sigma_c$ is less than 2 MeV \cite{Tanabashi:2018oca}. 
Furthermore $\Sigma_c^\ast$ decays to $\Lambda_c \pi$.  
Therefore, it may be difficult to observe the $\Sigma_c^\ast \bar D^{(\ast)}$
resonance states in the $J/\psi p$ channel. 
One may have to look into the $J/\psi p \pi$ channel to observe the pentaquark states consisting
mostly of the $\Sigma_c^\ast \bar D^{(\ast)}$ components.



{
In addition to the seven $\Sigma^{(\ast)}_c\bar{D}^{(\ast)}$ 
states predicted in the present study, 
we obtained two more narrow resonance states below the $\Lambda_c \bar D^\ast$ 
threshold in Ref.~\cite{Yamaguchi:2017zmn}, while such resonances have not been 
reported in experiments yet.
Although the OPEP does not contribute to the diagonal terms of the $\Lambda_c \bar D^\ast$ potential,
the $\Lambda_c\bar{D}^{(\ast)}-\Sigma^{(\ast)}_c\bar{D}^{(\ast)}$ coupling induces the OPEP and the attraction from the tensor term is also produced 
in the $\Lambda_c \bar D^{(\ast)}$ channel.
In future experiments, 
it is interesting to search for such resonances below the $\Lambda_c \bar D^\ast$ threshold.
}

\indent In conclusion,  by coupling the open charm meson-baryon channels to
a compact $uud c\bar{c}$ core with an interaction satisfying the heavy quark and chiral symmetries,
we predict the masses and decay widths of the three new pentaquark
states reported in  \cite{Aaij:2019vzc}.
Both the masses and widths of these three hidden-charm pentaquark states we have obtained are in 
reasonable agreement with the experimental results.
We point out that the three pentaquark states have quantum numbers  $J^P_{P_c^+(4312)} = 1/2^{-}$, $J^P_{P_c^+(4440)} = 3/2^{-}$, and $J^P_{P_c^+(4457)^{+}}= 1/2^{-}$
and the dominant molecular component of $P_c^+(4312)$ is the $\Sigma_c \bar D$ and that of $P_c^+(4440)$ and $P_c^+(4457)$ is 
$\Sigma_c \bar D^\ast$.
We find that the short range interaction by the coupling to the 5-quark-core states plays a major role in determining of the ordering of the multiplet states, while the long range force of the pion tensor force does 
in producing the decay widths,
which are consistent 
with the data. 
%
The importance of what we referred to as the chiral tensor dynamics is a universal feature for the heavy hadrons with light quarks.  
Such dynamical studies in coupled channel problems should be properly performed for further understanding of heavy hadron systems.  

The authors thank T. J. Burns for useful comments.
This work is supported in part  by the Special Postdoctoral Researcher  (SPDR) and iTHEMS Programs of RIKEN (Y.Y.) and by JSPS KAKENHI No.\ 16K05361 (C) (S.T.\ and M.T.), 
No. JP17K05441 (C),  and Grants-in-Aid for Scientific Research on Innovative Areas (No. 18H05407) (A.H.).
This work is also supported in part by the ``RCNP Collaboration Research network (COREnet)''.  


\bibliography{./reference} 
  
\end{document}